# Production of HD Molecules in Definite Hyperfine Substates

R. Engels,* K. Grigoryev, C. S. Kannis, Y. Michael, H. Ströher, and V. Verhoeven
*Institute for Nuclear Physics, Forschungszentrum Jülich, 52425 Jülich, Germany*

M. Büscher
*Peter Grünberg Institute, Forschungszentrum Jülich, 52425 Jülich, Germany and
Institute for Laser- and Plasma-Physics, Heinrich-Heine Universität Düsseldorf, 40225 Düsseldorf, Germany*

L. Huxold
*Institute for Laser- and Plasma-Physics, Heinrich-Heine Universität Düsseldorf, 40225 Düsseldorf, Germany*

L. Kochenda, P. Kravtsov, V. Trofimov, A. Vasilyev, and M. Vznuzdaev
*'Kurchatov Institute' B. P. Konstantinov Petersburg Nuclear Physics Institute (PNPI), 188300 Gatchina, Russia*
(Dated: January 27, 2020)

Polarized atomic beam sources have been in operation since many years to produce either nuclear polarized atomic hydrogen or deuterium beams. In recent experiments such a source was used to polarize both isotopes independently at the same time. By recombination of the atoms, HD molecules with all possible nuclear spin combinations can be created. Those spin isomers are useful for further applications like precision spectroscopy, as polarized targets for laser-particle acceleration, polarized fuel for fusion reactors or as an option for future measurements of electric dipole moments.

### INTRODUCTION

The Rabi apparatus [1] is the basis to build modern polarized atomic beam sources (ABS). Their development started in 1956 [2] and has led to a well-established technology. These ABS's produce a beam of hydrogen or deuterium atoms in dedicated hyperfine substates (HFS) like shown in table I. Experiments at storage rings use an ABS either to produce polarized ions for further acceleration [3] or to feed polarized internal storage-cell targets [4–8].

During the last years it was observed that the polarized hydrogen/deuterium atoms in such storage-cell targets can recombine into molecules where the nuclear polarization is partially preserved [9–11]. We built a dedicated apparatus for further investigations of the different recombination processes on various surfaces. With this setup it was possible to produce polarized $H_2$ and $D_2$ molecules with parallel nuclear spins. Depending on the surface materials large polarization values above 0.8 have been obtained [12].

Hydrogen molecules are a combination of fermions and exist as ortho-hydrogen, where both nuclear spins are coupled to a total spin $F = I_1 + I_2 = 1$, and para-hydrogen with $F = 0$. Ortho-hydrogen has a symmetric spin wave function and an anti-symmetric contribution due to the rotational magnetic moment $J$, which must be odd. The $F = 1$ state has three substates with $m_F = +1, 0$ and $-1$. Para-hydrogen with an anti-symmetric wave function due to anti-parallel proton spins must have even $J = 0, 2, 4, \ldots$ . The substates with $m_F = +1$ or $m_F = -1$ are called 'pure' states, because the proton spins of both atoms in the molecule are well defined, i.e. they must be parallel. The other states with $m_F = 0$ are 'mixed' states, either with symmetric (ortho) or anti-symmetric (para) combinations of the anti-parallel substates. Only the pure states are populated when polarized atoms with a defined nuclear spin from an ABS recombine into molecules.

Deuterons are bosons with $I = 1$. Thus, the total wave function of a $D_2$ molecules must be symmetric. If both spins are parallel and coupled to a total spin $F = 2$, the spin wave function is symmetric and the rotational moment $J$ must be even. The 9 possible combinations of the nuclear spins are divided into six ortho- and three para-deuterium substates. Again, only the substates with $m_F = +2$ or $-2$ are produced during the recombination of polarized atoms. Contrary to hydrogen it might be possible to freeze out these molecules as deuterium ice with $J = 0$.

TABLE I. The definition of the different hyperfine substates in the Zeeman region, described with $F = J + I$, and in the Paschen-Back region, described with $I$ and $J$. For hydrogen atoms in the $1S$ ground state the critical field amounts to $B_c(H) = 50.7$ mT and for deuterium to $B_c(D) = 11.4$ mT.

| Isotope | HFS | $B \ll B_c$ $\|F, m_F\rangle$ | $B \gg B_c$ $\|m_J, m_I\rangle$ |
|---|---|---|---|
| H | $\|H1\rangle$ | $\|1, 1\rangle$ | $\|1/2, 1/2\rangle$ |
|   | $\|H2\rangle$ | $\|1, 0\rangle$ | $\|1/2, -1/2\rangle$ |
|   | $\|H3\rangle$ | $\|1, -1\rangle$ | $\|-1/2, -1/2\rangle$ |
|   | $\|H4\rangle$ | $\|0, 0\rangle$ | $\|-1/2, 1/2\rangle$ |
| D | $\|D1\rangle$ | $\|3/2, 3/2\rangle$ | $\|1/2, 1\rangle$ |
|   | $\|D2\rangle$ | $\|3/2, 1/2\rangle$ | $\|1/2, 0\rangle$ |
|   | $\|D3\rangle$ | $\|3/2, -1/2\rangle$ | $\|1/2, -1\rangle$ |
|   | $\|D4\rangle$ | $\|3/2, -3/2\rangle$ | $\|-1/2, -1\rangle$ |
|   | $\|D5\rangle$ | $\|1/2, -1/2\rangle$ | $\|-1/2, 0\rangle$ |
|   | $\|D6\rangle$ | $\|1/2, +1/2\rangle$ | $\|-1/2, 1\rangle$ |

TABLE II. The hyperfine substates of HD molecules described with the total nuclear spin $F = I_p + I_d$ and its projection $m_F$ along an external magnetic field or directly with the single nuclear spins $I$ and their projections $m_I$.

| Isotope | HFS | $|F, m_F\rangle$ | $|m_{I_p}, m_{I_d}\rangle$ |
|---------|-----|------------------|----------------------------|
| HD      | $|HD1\rangle$ | $|3/2, 3/2\rangle$ | $|1/2, 1\rangle$ |
|         | $|HD2\rangle$ | $|3/2, 1/2\rangle$ | $|1/2, 0\rangle$ |
|         | $|HD3\rangle$ | $|3/2, -1/2\rangle$ | $|-1/2, 0\rangle$ |
|         | $|HD4\rangle$ | $|3/2, -3/2\rangle$ | $|-1/2, -1\rangle$ |
|         | $|HD5\rangle$ | $|1/2, 1/2\rangle$ | $|-1/2, 1\rangle$ |
|         | $|HD6\rangle$ | $|1/2, -1/2\rangle$ | $|1/2, -1\rangle$ |

For an HD molecule, a combination of a boson and a fermion, these selection rules do not apply. The possible substates are shown in table II. Here, every substate is a pure state and can have any rotational magnetic moment.

Up to now polarized solid HD ice has been produced with the 'brute force' method and was used as polarized target, e.g. at the CLAS experiment [13]. One advantage of solid HD is the long lifetime of the nuclear polarization which is in the order of years. But the polarization values achieved for the proton ($P_z \approx 0.5$) and the deuteron ($P_z \approx 0.27$) are rather small. Moreover, a control of the tensor-polarization $P_{zz}$ is not possible. For many important fundamental applications the other spin combinations are mandatory and larger polarization values would be helpful. Here we present a new method to produce all possible spin combinations of the proton and the deuteron in the HD molecule.

If an ABS is operated with a hydrogen and deuterium mixture, a beam containing the corresponding atoms in defined hyperfine substates can be generated. When these atoms recombine into HD molecules it is possible to produce any spin isomer of the HD molecules, because the nuclear spins of the hydrogen and the deuterium atoms can be adjusted separately.

## THE APPARATUS

Our ABS [5] consists of a dissociator in combination with a cooled nozzle to form an atomic hydrogen beam from molecular gas. After shaping the beam with two apertures, cylindrical sextupole magnets are used to focus the atoms with electron spin $m_J = +1/2$ on the beam axis and to defocus the others (Stern-Gerlach effect). A so-called medium magnetic field transition unit (MFT) to induce magnetic dipole $\pi$-transitions is utilized between the sextupole magnets to exchange the occupation number of the hyperfine substates $|H2\rangle \rightarrow |H3\rangle$, $|H1\rangle \rightarrow |H3\rangle$, $|D3\rangle \rightarrow |D4\rangle$ or $|D1\rangle \rightarrow |D4\rangle$. The following sextupole magnets defocus state $|H3\rangle$ and $|D4\rangle$. A weak field transition unit (WFT, again a $\pi$-transition) and a strong field transition unit (SFT; magnetic $\sigma$-transition), placed after the last sextupole magnet, can induce further transitions that are listed in table III. The necessary settings for the rf-frequency and the magnetic field of the WFT are very similar for hydrogen and deuterium, thus it can be used to induce the transitions for both isotopes simultaneously.

When these atoms are fed into a T-shaped storage cell at temperatures between 40 and 120 K, they can recombine into molecules (see Fig. 1). The ratio of the flux of hydrogen and deuterium into the ABS was tuned to be 2/3 to obtain an equal amount of hydrogen and deuterium atoms from the ABS in the recombination chamber. In this case the maximum amount of HD molecules and roughly equal amounts of $H_2$ and $D_2$ molecules are produced.

A longitudinal magnetic field along the cell, induced by a superconducting solenoid, increases the lifetime and the nuclear polarization is preserved for several ms, while the molecules stay inside the storage cell. This magnetic field couples to the nuclear spins and, therefore, the spin-spin coupling of the nucleons itself and their coupling with the rotational magnetic moment of the molecule $J$ are suppressed, which are the reasons for depolarization during wall collisions [12]. Up to 50% of the atomic polarization is found in these molecules after the recombination on a gold surface and a magnetic field higher than 0.3 T. A Fomblin surface allows for full vector-polarization preservation of Deuterium atoms at 0.5 T. An electron beam is then send through the cell to ionize the atoms and molecules. The corresponding ions, i.e. protons, deuterons, $H_2^+$, $D_2^+$ and $HD^+$, are accelerated to an energy of 1 to 2 keV by a positive potential on the gold surface [12].

All ions with the same kinetic energy reach the first component of the Lamb-shift polarimeter (LSP), i.e. a Wien filter that is used in two different ways: It is able to separate the different velocities, i.e. the different ions according to their masses. Only deuterons and $H_2^+$ ions, both with mass two, cannot be distinguished. At the same time the transversal magnetic field induces a Larmor precession of the longitudinally oriented nuclear spin. If the magnetic field is chosen in a way that the time-of-flight of the ions through the Wien filter allows a spin rotation of 180°, the orientation of the spin direction is changed, but not its projection onto the beam axis. The coupling of the nuclear spins to the spin of the single electron of the molecular ions is responsible for an adiabatic realignment along the radial magnetic field. Afterwards, the ions undergo a charge exchange with Cesium vapor to form metastable atoms in the first excited state 2S. A strong magnetic field along this Cesium cell reorientates the spins of the molecular ions back on the beam direction and preserves the nuclear polarization until the field is much larger than the critical field of these metastable atoms ($B_c = 6.34$ mT for hydrogen and $B_c = 1.54$ mT for deuterium). With a longitudinal magnetic field around 57 mT, a radial electric field of about 10 V/cm and a radio frequency of 1.60975 GHz inside a cavity in the $TM_{010}$ mode, the spinfilter is able to quench all metastable atoms into the ground state. Only at special resonances at different magnetic fields metastable atoms in single HFS can pass through. In the last section these excited atoms are quenched into the ground state with a strong electric field (Stark effect) and the produced Lyman-$\alpha$ photons are registered with a photomultiplier as function of the magnetic field in the spinfilter. The amount of measured photons $N_+$ and $N_-$ at the resonances corresponds linearly to the number of incoming ions with the same nuclear spin. Therefore, the nuclear vector-polarization $P_z$ of the protons/deuterons can be deduced by comparing the photon numbers $P_z = \frac{N_+ - N_-}{N_+ + N_-}$ [14, 15]. The resonances of the HFS of the metastable hydrogen and deuterium are well separated. Thus, even when the Wien filter is not able to separate $H_2^+$ ions and deuterons, or when $HD^+$ molecules will reach the Cesium cell, the LSP can measure the nuclear polarization of both nucleons independently.

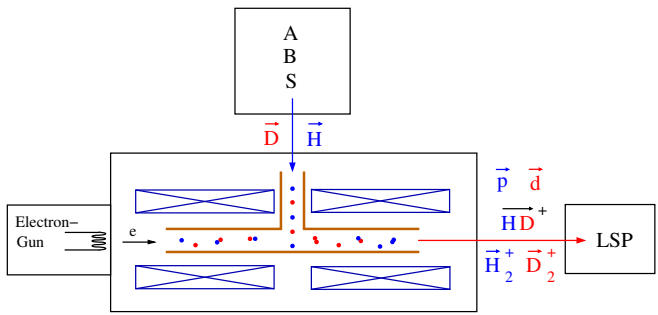

FIG. 1. Experimental setup taken from Ref. [12] with minor modifications: A storage cell inside a superconducting magnet is fed by the polarized H and D beam from the atomic beam source (ABS). Atoms and molecules from recombination are ionized by electron impact. The cell is set to a potential up to +2 keV to accelerate the protons/deuterons and molecular ions into the Lamb-shift polarimeter (LSP), which allows one to measure the polarization.

## EXPERIMENTAL RESULTS

The standard settings of our ABS allow only the production of the spin isomer $|HD4\rangle$, i.e. proton and deuteron spin are anti-parallel to the quantization axis (see table III). In this case, the first set of sextupole magnets will defocus the hydrogen atoms in the HFS $|H3\rangle$ and $|H4\rangle$ and the deuterium atoms in the substates $|D4\rangle$, $|D5\rangle$ and $|D6\rangle$. The MFT is used to transfer deuterium atoms from HFS $|D3\rangle$ into $|D4\rangle$. The next set of sextupole magnets defocuses the deuterium atoms in state $|D4\rangle$ again, so that only the HFS $|H1\rangle$, $|H2\rangle$, $|D1\rangle$ and $|D2\rangle$ are left in the beam. Now, the WFT will induce the corresponding transitions for hydrogen and deuterium at the same time ($|H1\rangle + |H2\rangle \to |H2\rangle + |H3\rangle$ and $|D1\rangle + |D2\rangle \to |D3\rangle + |D4\rangle$). Thus, the beam includes mostly atoms with anti-parallel nuclear spins, i.e. the HFS $|H2\rangle + |H3\rangle$ for hydrogen and $|D3\rangle + |D4\rangle$ for deuterium. These atoms recombine on a Fomblin (Perfluroether) surface into $H_2$, $D_2$ and HD molecules. After ionization by electron impact, acceleration of the molecular ions into the LSP and mass separation by the Wien filter, the polarization of the $HD^+$ ions is deduced from the Lyman-$\alpha$ spectra (see Fig. 2). The polarization can be deduced from the peak ratios to $P_z = -0.77 \pm 0.01$ for hydrogen and $P_z = -0.79 \pm 0.01$ and $P_{zz} = +0.69 \pm 0.01$ for deuterium.

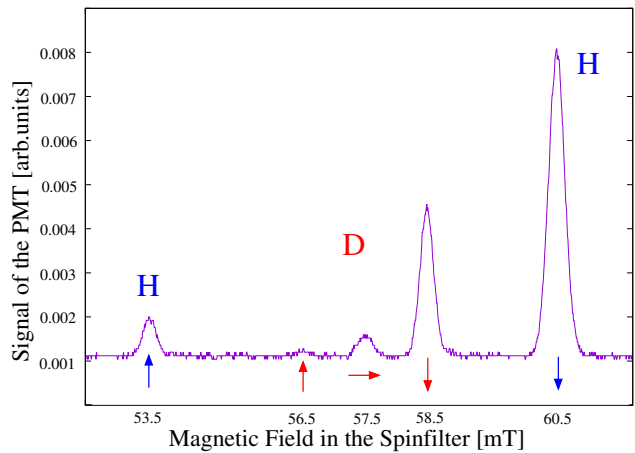

FIG. 2. Lyman-$\alpha$ spectrum of HD molecules in the spin isomer $|HD4\rangle$ after recombination on a Fomblin surface. The nuclear polarization deduced from this spectra is $P_z = -0.77 \pm 0.01$ for hydrogen and $P_z = -0.79 \pm 0.01$ and $P_{zz} = +0.69 \pm 0.01$ for deuterium.

The necessary settings for the production of the other spin isomers of HD molecules are listed in table III. With the simltaneous use of a WFT and a SFT or two SFT all possible spin directions of the proton and the deuteron are possible and in range after minor modifications of the existing ABS.

TABLE III. The settings of the ABS transition units to get an atomic beam of hydrogen and deuterium with dedicated nuclear spins to produce all different spin isomers of HD molecules.

| 1. Transition unit | 2./3. Transition unit | Atomic beam | HFS of HD molecules |
|---|---|---|---|
| MFT: $|D3\rangle \to |D4\rangle$ | SFT: $|H2\rangle \to |H4\rangle$ <br> SFT: $|D2\rangle \to |D6\rangle$ | $|H1\rangle + |H4\rangle$ <br> $|D1\rangle + |D6\rangle$ | $|HD1\rangle$ |
| MFT: $|D1\rangle \to |D4\rangle$ | SFT: $|H2\rangle \to |H4\rangle$ <br> SFT: $|D3\rangle \to |D5\rangle$ | $|H1\rangle + |H4\rangle$ <br> $|D2\rangle + |D5\rangle$ | $|HD2\rangle$ |
| MFT: $|D1\rangle \to |D4\rangle$ | WFT: $|H1\rangle + |H2\rangle \to |H2\rangle + |H3\rangle$ <br> SFT: $|D3\rangle \to |D5\rangle$ | $|H2\rangle + |H3\rangle$ <br> $|D2\rangle + |D5\rangle$ | $|HD3\rangle$ |
| MFT: $|D3\rangle \to |D4\rangle$ | WFT: $|H1\rangle + |H2\rangle \to |H2\rangle + |H3\rangle$ <br> WFT: $|D1\rangle + |D2\rangle \to |D3\rangle + |D4\rangle$ | $|H2\rangle + |H3\rangle$ <br> $|D3\rangle + |D4\rangle$ | $|HD4\rangle$ |
| MFT: $|D3\rangle \to |D4\rangle$ | WFT: $|H1\rangle + |H2\rangle \to |H2\rangle + |H3\rangle$ <br> SFT: $|D2\rangle \to |D6\rangle$ | $|H2\rangle + |H3\rangle$ <br> $|D1\rangle + |D6\rangle$ | $|HD5\rangle$ |
| MFT: $|D3\rangle \to |D4\rangle$ | SFT: $|H2\rangle \to |H4\rangle$ <br> WFT: $|D1\rangle + |D2\rangle \to |D3\rangle + |D4\rangle$ | $|H1\rangle + |H4\rangle$ <br> $|D3\rangle + |D4\rangle$ | $|HD6\rangle$ |

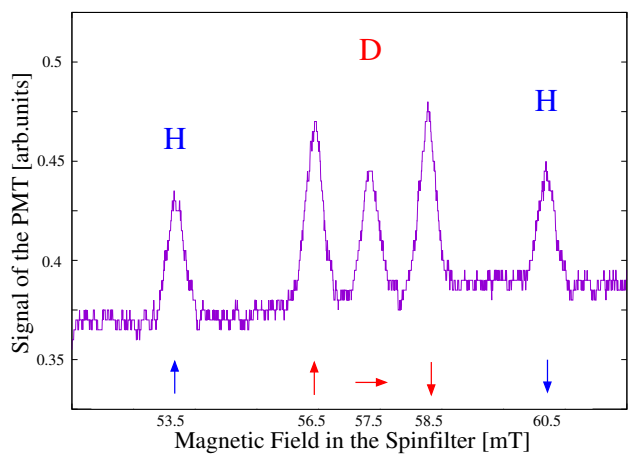

FIG. 3. Lyman-$\alpha$ spectrum of HD molecules when unpolarized hydrogen atoms recombine on a gold surface mostly with deuterium atoms in the HFS $|D3\rangle$ or $|D6\rangle$. The protons are unpolarized and the deuterons with $m_I = 0$ are suppressed. The observed tensor-polarization of $P_{zz} = +0.22 \pm 0.02$ is rather low due to some technical problems with one transition unit of the ABS during these measurements.

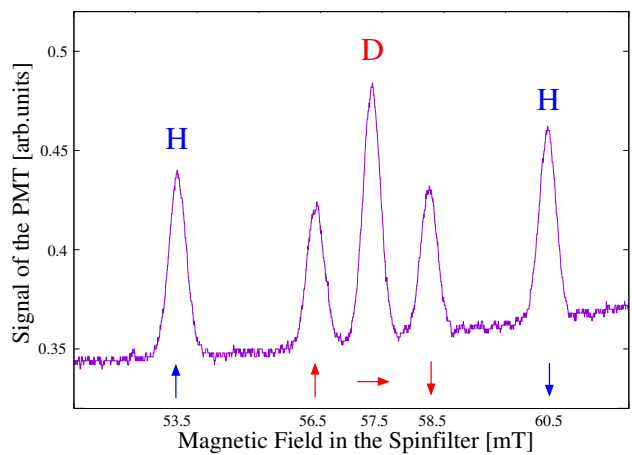

FIG. 4. Lyman-$\alpha$ spectrum of HD molecules when unpolarized hydrogen atoms recombine on a gold surface mostly with deuterium atoms in $|D2\rangle$ and $|D5\rangle$. The protons are unpolarized and the deuteron spin is perpendicular to the quantization axis with a tensor-polarization of $P_{zz} = -0.61 \pm 0.01$.

Instead of Fomblin most measurements were carried out on a gold surface. As expected from Ref. [12], Fomblin allows twice larger polarization values for the molecules, because this surface preserves the nuclear polarization during the recombination process. Nevertheless, most measurements are made with a gold surface, because the use of Fomblin oil pollutes all surfaces in the vacuum chamber and reduces the pumping speed of the internal cryo-panels by orders of magnitude.

When the MFT transfers HFS $|D1\rangle$ into $|D4\rangle$ and the SFT is used at the end, it is possible to produce only tensor-polarized deuterium. If the SFT transfers $|D2\rangle \to |D6\rangle$ ($\sigma$ transition) the deuterium atoms are in the HFS $|D3\rangle$ and $|D6\rangle$. Recombination with unpolarized hydrogen atoms gives equal amounts of the substates $|HD1\rangle$, $|HD4\rangle$, $|HD5\rangle$, and $|HD6\rangle$, i.e. only states with positive tensor-polarization of the deuteron (see Fig. 3). Negative tensor-polarization can be produced if the SFT $|D3\rangle \to |D5\rangle$ $\sigma$-transition is induced to get $|D2\rangle$ and $|D5\rangle$. These atoms can recombine with unpolarized hydrogen to produce the states $|HD2\rangle$ and $|HD3\rangle$ with equal amounts (Fig. 4).

# DISCUSSION AND OUTLOOK

As in previous works on the production of polarized $H_2$ and $D_2$ molecules due to the recombination of polarized H or D atoms, the creation of polarized HD molecules in single spin isomers was investigated. Depending on the spin states of the different isotopes, defined with the corresponding settings of the ABS, all possible combinations of the proton and deuteron spin are accessible with an ABS.

In former polarization measurements of the $H_2$ and $D_2$ molecules recombined on a gold surface polarization values up to $P_z = 0.45$ were observed, about 50% of the atomic polarization [12]. During these measurements the polarization of the $H_2$, the $D_2$ and the nucleons in the HD molecules does not exceed 0.35. The reason is still not completely understood, but seems to depend on the production method of the self-made gold cells that had to be changed recently.

Compared to the measurements with protons/deuterons at the entrance of the Cesium-cell the molecular ions have more options to interact with the Cesium. For example, the stripping process into two protons or the charge exchange into fast molecules are responsible for a larger background in the Lyman-$\alpha$ spectra. The interaction of the ions in the beam with the magnetic field of the spin-filter is responsible for the increase of the background at higher magnetic fields.

In the next step the polarized $D_2$ and HD molecules will be frozen as polarized ice on a cold surface inside a strong magnetic field exploiting the higher evaporation temperature of the $H_2$ molecules. The goal is to build polarized targets for laser-acceleration experiments to produce polarized proton and deuteron beams [16].

Polarized $H_2^+$ and $HD^+$ ions in single spin-isomers are a perfect tool for precision spectroscopy of this most simple molecular ions, because the influence of the spin-dependent part of the transition-frequencies can be determined separately [17].

Another option to use these molecules can be the search for a static electric dipole moment (EDM) with oscillating electric fields [18] or even experiments searching for an oscillating nuclear EDM produced by the axion dark matter field [19]. For both methods the use of highly polarized HD molecules is beneficial, because the EDM effect would be proportional to the inverse nucleon charge $z^{-1}$ that is at minimum for the hydrogen isotopes.

Another promising application of nuclear polarized hydrogen molecules and its isotopes would be the use as polarized fuel for fusion reactors [20]. Besides $D_2$ molecules, polarized DT would be needed, either with parallel nuclear spins (Fig. 2) to increase the reaction rate of the fusion reactions or with only tensor-polarized deuterons (Fig. 4) to focus the neutrons on special blanket areas, e.g. to increase the lifetime of the blanket or to breed new tritium via the $^6Li + n \rightarrow\ ^4He + t$ reaction. To avoid the radioactive tritium, the HD molecules are a perfect training ground to study the production, the storage and the handling of these molecules for further use, because the hyperfine structures of HD and DT are almost identical.

The authors wish to thank the ATHENA project of the Helmholtz Association (HGF) for financial support. L. Huxold acknowledges support by the Deutsche Forschungsgemeinschaft (BU 2227/1-1).